\documentclass[twocolumn,prl,aps,showpacs,preprintnumbers,superscriptaddress]{revtex4}

\usepackage{graphicx}
\usepackage{amsmath}

\newcommand{\be}{\begin{equation}}
\newcommand{\ee}{\end{equation}}
\newcommand{\bea}{\begin{eqnarray}}
\newcommand{\eea}{\end{eqnarray}}
\newcommand{\HH}{{\cal H}}

\newcommand{\p}{\partial}

\newcommand{\la}{\langle}
\newcommand{\ra}{\rangle}

\newcommand{\lp}{\left(}
\newcommand{\rp}{\right)}

\renewcommand{\epsilon}{\varepsilon}

\begin{document}
\title{Klein Backscattering and Fabry-P\'erot Interference 
in Graphene Heterojunctions}

\author{Andrei V. Shytov}
\affiliation{Department of Physics,
University of Utah,
Salt Lake City, UT 84112}
\author{Mark S. Rudner}
\affiliation{Department of Physics, Massachusetts Institute of Technology, Cambridge MA 02139}
\author{Leonid S. Levitov}
\affiliation{Department of Physics, Massachusetts Institute of Technology, Cambridge MA 02139}


\begin{abstract}
We present a theory of quantum-coherent transport through a lateral p-n-p structure in graphene, which fully accounts for the interference of forward and backward scattering on the p-n interfaces. 
The backreflection amplitude changes sign at zero incidence angle because of the Klein phenomenon, adding a phase $\pi$ to the interference fringes. 
The contributions of the two p-n interfaces to the phase of the interference cancel with each other at zero magnetic field, but become imbalanced at a finite field. The resulting half-period shift in the Fabry-P\'erot fringe pattern, induced by a relatively weak magnetic field, can provide a clear signature 
of Klein scattering in graphene. This effect is shown to be robust in the presence of spatially inhomogeneous potential of moderate strength.
\end{abstract}
\maketitle
\vspace{-10mm}

The electron system in graphene features surprising connections with the
relativistic quantum mechanics of a massless Dirac particle in external electric and magnetic fields \cite{Gusynin05,Peres06,Katsnelson06,Cheianov06,Shytov07}. 
In particular, charge flow in lateral p-n junctions in graphene is described by Klein scattering \cite{Katsnelson06,Cheianov06,Beenakker07}, exhibiting perfect transmission through the barrier at normal incidence, regardless of the barrier characteristics, and a barrier-dependent, finite reflection coefficient at non-normal incidence \cite{Klein29}. On the experimental side, while transport properties of the first p-n junctions fabricated in graphene were dominated by disorder \cite{Huard07,Williams07,Ozyilmaz07},
two recent papers \cite{Gorbachev08,Stander08} report on observations of a contribution to the conductance consistent with expectations for ballistic transmission through p-n interfaces. To extract this ``intrinsic'' contribution, which is relatively small in magnitude, one must account for the screening of the gate potential \cite{Fogler08a} and the effects of disorder \cite{Fogler08b}.

\begin{figure}
\includegraphics[width=3.3in]{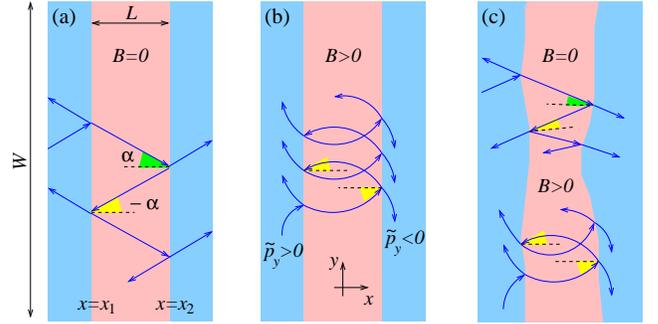}
\caption[t]{Schematic of electron transmission through a p-n-p structure at zero (a) and finite (b) magnetic field $B$. The amplitude of backeflection changes sign at the incidence angle $\alpha=0$. At finite $B$, for the trajectories 
satisfying condition (\ref{eq:py_range}) the angles of incidence on both p-n interfaces are of the same sign. This adds a phase $\pi$ the electron phase accumulated between reflections, resulting in half a period shift of Fabry-P\'erot interference fringes. (c) Weak spatial inhomogeneity does not alter the relative sign of incidence angles.
}
 \label{fig0}
\vspace{-5mm}
\end{figure}

What other features, besides collimated transmission, may serve as an experimental signature of Klein scattering?
Here we focus on the characteristic behavior of the reflection amplitude, which exhibits a jump in phase by $\pi$ when the incidence angle $\alpha$ is varied from positive to negative values (see Fig.\ref{fig0}). The 
sign change occurs at normal incidence
because the reflection amplitude
vanishes at $\alpha=0$. 
Below we show that this phase shift, which is fundamental to Klein scattering, could serve as a hallmark of Klein physics in graphene.


The backreflection phase can be detected from interference of electron waves scattered on two parallel p-n boundaries in a p-n-p structure. Transmission in this system, described by the Fabry-P\'erot (FP) model, exhibits periodic dependence on the phase $\Delta\theta$ gained by an electron bouncing between the p-n interfaces (see Fig.\ref{fig0}),
%
\be\label{eq:theta_total}
\Delta\theta = 2\theta_{\rm WKB}+\Delta\theta_1+\Delta\theta_2
,
\ee
where $\theta_{\rm WKB}=\frac1{\hbar}\int_1^2p_x(x')dx'$ is the WKB phase and $\Delta\theta_{1(2)}$ are the backreflection phases for the interfaces 1 and 2,
exhibiting a $\pi$-jump at zero incidence angle $\alpha$.


As illustrated in Fig.\ref{fig0}, the contribution $\Delta\theta_1+\Delta\theta_2$ to the net phase can be altered by a magnetic field. At zero $B$ the incidence angles at interfaces 1 and 2 have opposite signs,
and thus the jumps in $\Delta\theta_{1(2)}$ cancel.
However, for curved electron trajectories at a finite $B$, the signs of the incidence angles can be made 
equal.
Indeed, because of translational invariance along the p-n interface, in the presence of a magnetic field 
the $y$-component of electron kinetic momentum varies in space as 
$\tilde p_y(x)=p_{y}-eBx$,
where $p_{y}$ is the conserved canonical momentum component that labels different trajectories. For the incidence angles at interfaces 1 and 2 to be of equal sign, $\tilde p_y(x_1)$ and $\tilde p_y(x_2)$ must have opposite signs,
which happens when 
%
\be\label{eq:py_range}
-eBL/2<p_{y}<eBL/2
.
\ee
%
In this case the net backreflection phase $\Delta\theta_1+\Delta\theta_2$ in (\ref{eq:theta_total}) equals $\pi$. As we shall see, the backreflection phase manifests itself as \emph{half a period shift} of the FP fringe contrast. 
This phase shift develops for the field strength 
such that the range (\ref{eq:py_range}) exceeds the Klein collimation range
in which the p-n interface is transparent, $eB\gtrsim \Delta p/L$ (a similar condition appeared in the discussion of magnetoresistance in Ref.\cite{Cheianov06}). 

One useful characteristic of the phase jump in backreflection
is that it is less momentum-selective than collimated transmission. A potential difficulty, however, is that the interference of scattering on two p-n interfaces can be sensitive to disorder. Below we will investigate the dependence of the FP contrast on magnetic field in the presence of large-scale spatial fluctuations.
We find that, while 
the FP fringe contrast is suppressed, the 
$1/2$-period shift,
controled by
backreflection, remains surprisingly robust. Even at a relatively high disorder strength, when the FP contrast is strongly reduced and the fringes become aperiodic, the $1/2$-period shift induced by the magnetic field remains clearly discernable. 

Here we shall model the gate potential by a parabola $U(x)=ax^2-\epsilon$, which at $\epsilon>0$ creates p-n interfaces at
\be\label{eq:x_epsilon}
x =\pm x_\epsilon
,\quad
x_\epsilon \equiv \sqrt{\epsilon/a}
.
\ee
The energy $\epsilon$ is controled by a top gate, with $\epsilon=\alpha V_{\rm tg}$ (in Ref.\cite{Gorbachev08}, $\alpha\approx \frac1{300}$). The curvature parameter $a$ is determined by the width of the top gate $L_{\rm tg}$ and its height $h$ above the graphene plane. The actual potential profile may be nonparabolic (see Ref.\cite{Gorbachev08} for modelling of screening effects), however this difference should not 
matter for FP interference, occurring when the separation between p-n interfaces $L=2x_\epsilon$ spans only few de Broglie wavelengths. This is the case near the threshold in $V_{\rm tg}$ at which the p-n-p structure forms, where $L$ is small compared to $h$ and $L_{\rm tg}$. The parabolic $U(x)$ may provide a reasonable approximation also for the devices described in Ref.\cite{Stander08}, especially those with narrow top gates.

The Hamiltonian for a two-component Dirac wavefunction, in the presence of the potential $U(x)$, is
\be
\HH = v_F \sigma_3p_x + v_F \sigma_2 (p_{y}-eBx) + U(x)
\ee
where $\sigma_{2,3}$ are Pauli matrices, $v_F\approx 1.1\cdot 10^6\,{\rm m/s}$ is the Fermi velocity in graphene, and $B$ is the magnetic field. Hereafter we set $\hbar=v_F=1$, with the units for energy, length and magnetic field 
as follows:
\bea\nonumber
&&\epsilon_*=(av_F^2\hbar^2)^{1/3}\approx 14\,{\rm meV}
,\quad 
 x_*=\hbar v_F/\epsilon_*\approx 53\,{\rm nm}
,
\\ \label{eq:units}
&& B_*=\Phi_0/2\pi x_*^2\approx 0.24\,{\rm T}
,\qquad (\Phi_0=h/e)
,
\eea
where we used the curvature $a$ of $5\,{\rm eV/\mu m^2}$, obtained from the model potential of Ref.\cite{Gorbachev08} fitted to a parabola.
Because of the weak $a^{1/3}$ dependence, the estimates (\ref{eq:units}) should also apply, at least roughly, to other systems.

\begin{figure} 
\includegraphics[width=3.2in]{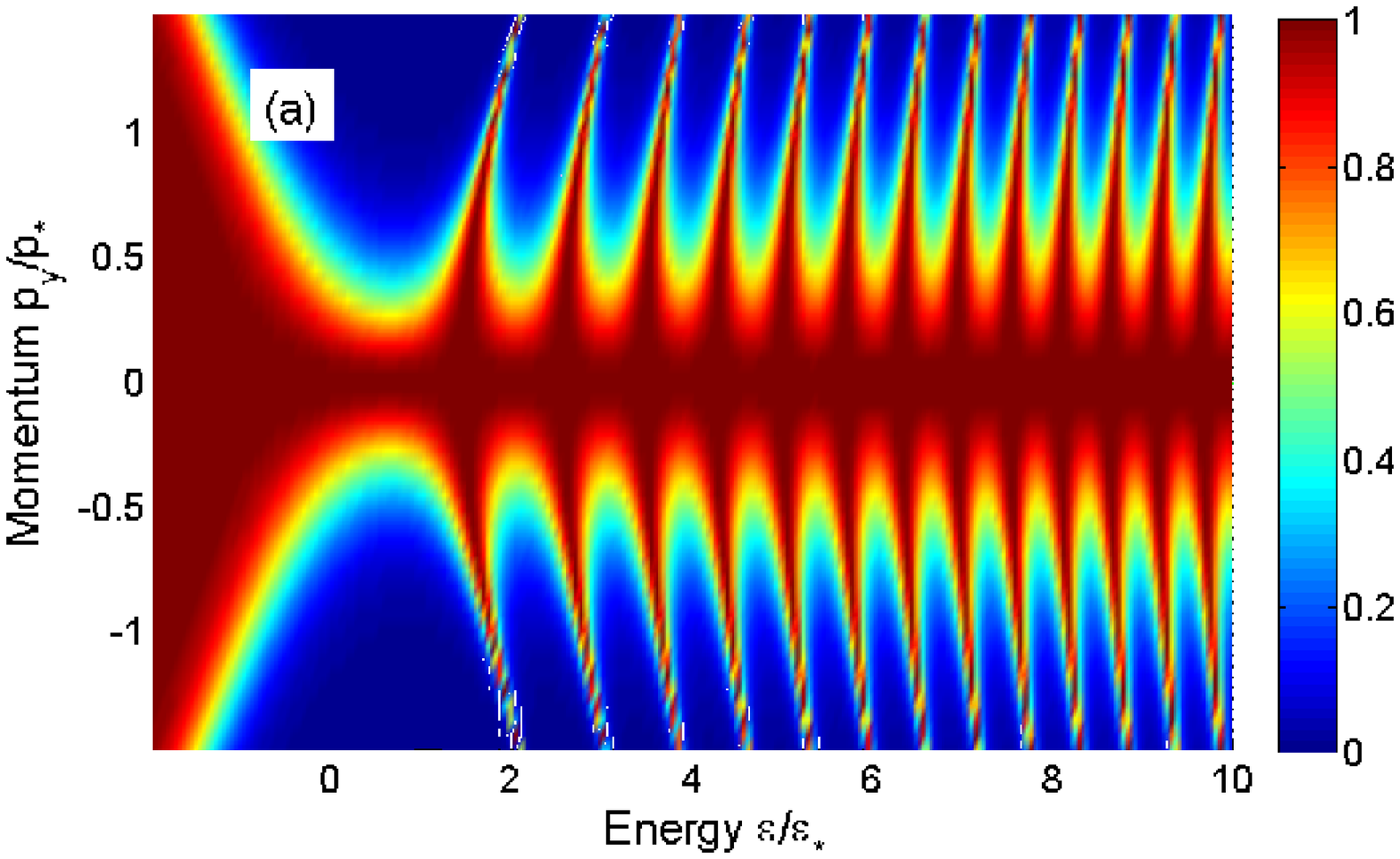}
\includegraphics[width=3.2in]{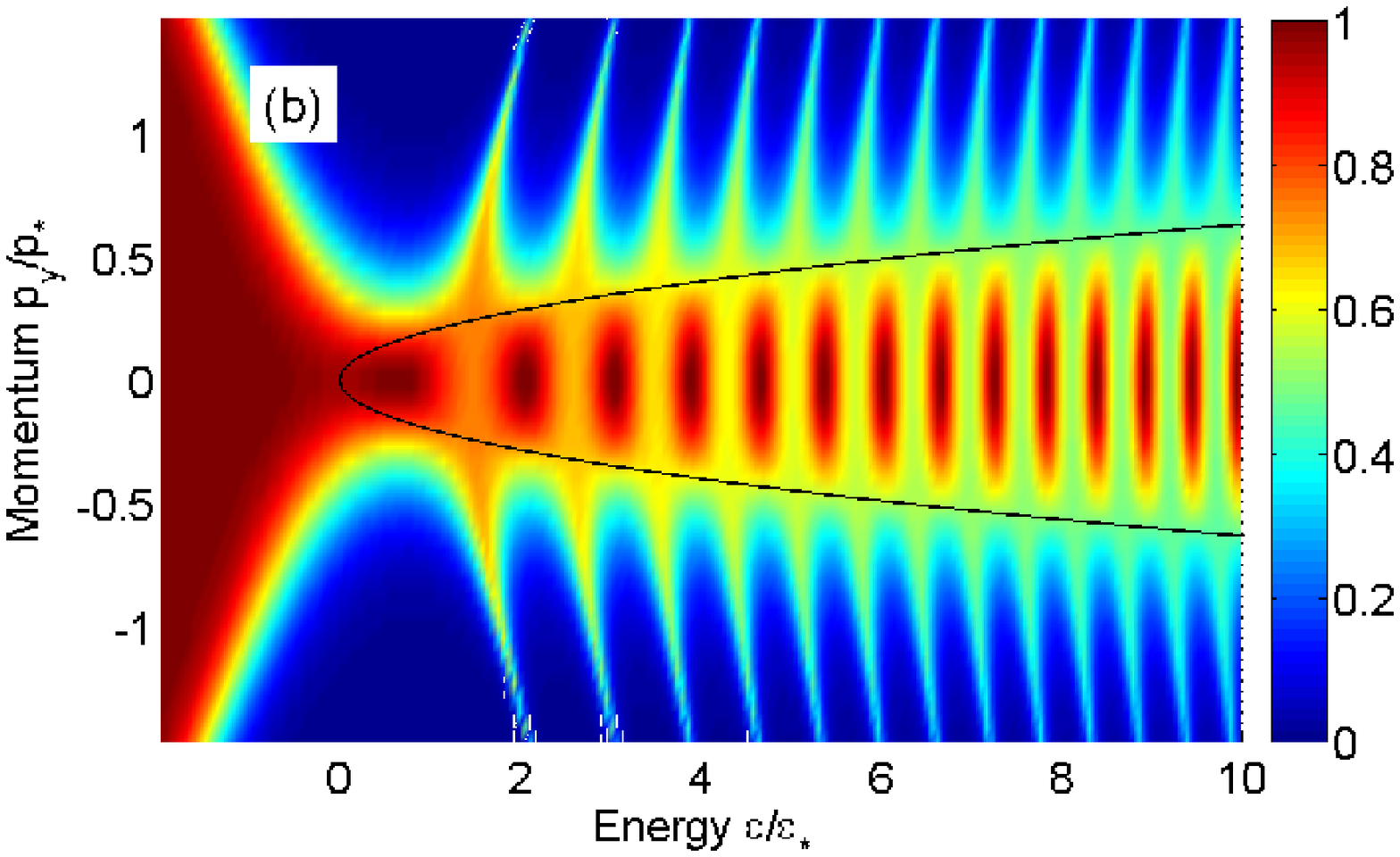}
\caption[t]{Transmission of the p-n-p structure (a) at zero magnetic field and (b) at a finite field of $B=0.2\Phi_0/x_*^2$. Electron momentum $p_y$ is measured in units of $p_*=v_F/\epsilon_*$. Note that at finite $B$ the fringe contrast is reversed across the parabola (\ref{eq:parabola}) (black line), with the maxima and minima of the fringe pattern interchanging. This behavior is in agreement with the Fabry-P\'erot model (\ref{eq:FPgeneral}), with the reflection amplitude vanishing on the parabola, and changing its sign.
}
 \label{fig1}
\vspace{-5mm}
\end{figure}

To find the transmission and reflection coefficients, we factorize the wavefunction as $\psi(x,y)=e^{ip_{y} y}\psi(x)$, and solve the one-dimensional Schr\"odinger equation 
\be\label{eq:LZdynamics}
i\p_x\psi=\lp U(x)\sigma_3-i(p_{y}-eBx)\sigma_1\rp\psi
,
\ee
where without loss of generality we set the Fermi energy equal to zero. Transmission, evaluated from numerical solution of Eq.(\ref{eq:LZdynamics}), exhibits resonances as a function of energy and momentum, shown in Fig.\ref{fig1}. We note a drastic difference between the results at zero $B$,
which are identical to those of Ref.\cite{Silvestrov07}, and 
the results at finite $B$.


To understand the behavior of transmission, it is instructive to view the differential equation (\ref{eq:LZdynamics})
as a fictitious time-dependent Schr\"odinger evolution of a two-level system, the coordinate $x$ playing the role of time. In this analogy, 
the system is driven through an avoided level crossing with splitting $-2i(p_{y}-eBx)$ at the crossing times determined by degeneracy of the diabatic states,
$U(x)= ax^2-\epsilon=0$. This condition yields $x=\pm x_\epsilon$, Eq.(\ref{eq:x_epsilon}).
A Landau-Zener transition at the first level crossing creates a coherent superposition of the diabatic states that can interfere at the second crossing. This so-called St\"{u}ckelberg inteference (see \cite{Berns08} and references therein), which can be constructive or destructive, is described by an oscillatory function of the phase $\Delta\theta=-2\int_{-x_\epsilon}^{x_\epsilon}U(x)dx=\frac43\epsilon x_\epsilon$ gained between the crossings. The locations of interference fringes, determined from the condition $\Delta\theta=2\pi n$ with $n=1,2...$, are $\epsilon_n=\lp 3\pi n/2\rp^{2/3}\epsilon_*$, which agrees with the positions of the fringes seen in Fig.\ref{fig1}.

For $B=0$, the suppression of oscillations near zero $p_{y}$ can be linked to the absence of Landau-Zener transitions at vanishing level-splitting. In the scattering picture, this is nothing else than the Klein phenomenon of perfect transmission at normal incidence. 
At finite $B$, the oscillations disappear when one of the level splittings $p_{y}\pm eBx_\epsilon$ vanishes, suppressing one of the Landau-Zener transitions. In terms of electron motion, this condition is equivalent to the requirement of normal incidence on either of the interfaces (\ref{eq:x_epsilon}), giving
\be\label{eq:parabola}
p_{y}=\pm eB \sqrt{\epsilon/a}
, 
\ee
which is the black parabola drawn in Fig.\ref{fig1}b. Indeed, fringes disappear on this line; 
upon crossing the line, 
the maxima and minima of fringes interchange, indicating a $\pi$ phase shift in the phase of fringe contrast. 

A more refined description can be obtained from a quasiclassical solution \cite{Silvestrov07} with position-dependent momentum 
$p_x(x)=\sqrt{U^2(x)-\tilde p_y^2(x)}$.
%
%
The turning points, defined by $p_x=0$, are arranged as
$x_{1'}<x_1<x_2<x_{2'}$, with $x_2$ and $x_{2'}$ equal
%
\be
\sqrt{\epsilon+b^2- p_{y}}+b
,\quad
\sqrt{\epsilon+b^2+ p_{y}}-b
,
\ee
where $b=\frac12eB$ and $x_{1(1')}(p_y)=-x_{2(2')}(-p_y)$. Hereafter we set $a=1$, restoring physical units later.
Remarkably, the conditions $x_1=x_{1'}$ and $x_2=x_{2'}$, which correspond to one of the p-n interfaces becoming transparent because of the Klein phenomenon, yield
a relation between $p_{y}$ and $\epsilon$ which is identical to Eq.(\ref{eq:parabola}) found above.

The classically forbidden regions $x_{1'}<x<x_1$ and $x_2<x<x_{2'}$, where $p_x$ is imaginary, correspond to the Klein barriers at the interfaces 1 and 2. Denoting the corresponding transmission coefficients as $t_1$ and $t_2$, we can write the net transmission of the entire p-n-p structure in a general Fabry-P\'erot form
\be\label{eq:FPgeneral}
T(\epsilon,p_{y}) = \frac{t_1t_2}{\left|1-\sqrt{r_1r_2}e^{i\Delta\theta}\right|^2}
\ee
where $r_{1(2)}=1-t_{1(2)}$ are the reflection coefficients,
and the phase $\Delta\theta$ is a sum of the WKB part and the phases of the reflection amplitudes,  Eq.(\ref{eq:theta_total}).

The transmissions $t_{1(2)}$ 
can be evaluated in the WKB tunneling approximation:
%
\be\label{eq:WKB_t1}
t_1\!\!=e^{-2{\rm Im}\int_{x_1}^{x_{1'}} \!\! p_x(x')dx'}\!\!\approx e^{-\lambda(p_{y}-eBx_\epsilon)^2}
\!\!,\quad
\lambda=\frac{\pi}{2ax_\epsilon}
,
\ee
with the integral computed by linearizing $U(x)$ near $x=x_\epsilon$. Similarly, linearizing $U(x)$ near $x=-x_\epsilon$, we find 
$t_2=e^{-\lambda(p_{y}+eBx_\epsilon)^2}$. 
Thus, the reflection amplitudes are
\be\label{eq:WKB_r1}
{\rm sgn}(p_{y}\pm eBx_\epsilon)\,e^{i\theta_{\rm reg}(p_y)}\,\sqrt{1-e^{-\lambda(p_{y}\pm eBx_\epsilon)^2}} 
,
\ee
where we factored the sign, responsible for the phase jump,
and a regular part of the phase $e^{i\theta_{\rm reg}}$, as follows from analyticity in $p_y$.
Because the WKB treatment is exact for linear potentials \cite{Shytov07b}, and the transmission (\ref{eq:WKB_t1}) is exponentially small unless $|p_{y}\pm eBx_\epsilon|\lesssim \lambda^{-1/2}$, the linearization of $U(x)$ used to evaluate the integral in (\ref{eq:WKB_t1}) gives accurate results for the energies of interest, $\epsilon\gtrsim\epsilon_*$.

\begin{figure}
\includegraphics[width=2.8in]{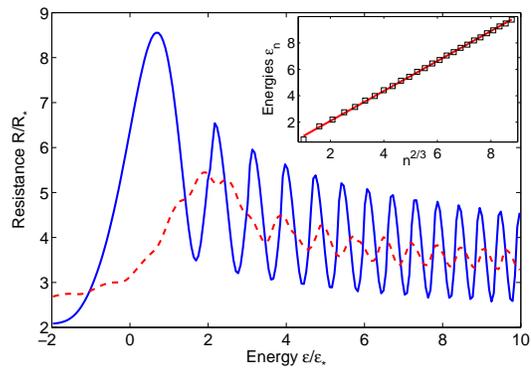}
\caption[t]{
Fringes in resistance (\ref{eq:resistance}) at $B=0$, plotted in the units of $R_*=(x_*/W)h/4e^2$ (blue line). Inset illustrates a $n^{2/3}$ scaling for the maxima and minima of $R$, which is consistent with the $\epsilon^{3/2}$ dependence of the WKB phase (\ref{eq:WKB_theta}).
Averaging over smooth potential fluctuations, described by a sum of a few harmonics, suppresses fringe contrast (red dashed line). 
Here we use Eq.(\ref{eq:T_average}) with $\sum_m|a_m|=3\epsilon_*$.
}
 \label{fig2}
\vspace{-5mm}
\end{figure}

The dispersion of the resonances in Fig.\ref{fig1} can be understood from the momentum dependence of the quasiclassical WKB phase 
(for simplicity, we set $B=0$):
\be\label{eq:WKB_theta}
\theta_{\rm WKB}=\int^{x_2}_{x_1} p_x(x')dx'
\approx \frac43\epsilon^{3/2}-\frac{p_{y}^2}{2\epsilon^{1/2}}\log\frac{\epsilon}{|p_{y}|}
,
\ee
where an expansion in the parameter $|p_{y}/\epsilon|\ll 1$ is legitimate because Klein collimation restricts transmission to $\Delta p_{y}\sim \lambda^{-1/2}$. 
The quantization condition $\theta_{12}=\pi (n+\frac12)$ gives the resonance energies $\epsilon_n(p_{y})$ dispersing as in Fig.\ref{fig1}. 

To summarize, the FP model (\ref{eq:FPgeneral}) 
is in full agreement with our numerical results. In particular, it explains the striking difference between the behavior at zero and finite $B$, as well as the phase shift of the fringe pattern, resulting from a sign change of the reflection amplitudes, 
Eq.(\ref{eq:WKB_r1}).


\begin{figure}
\includegraphics[width=3.2in]{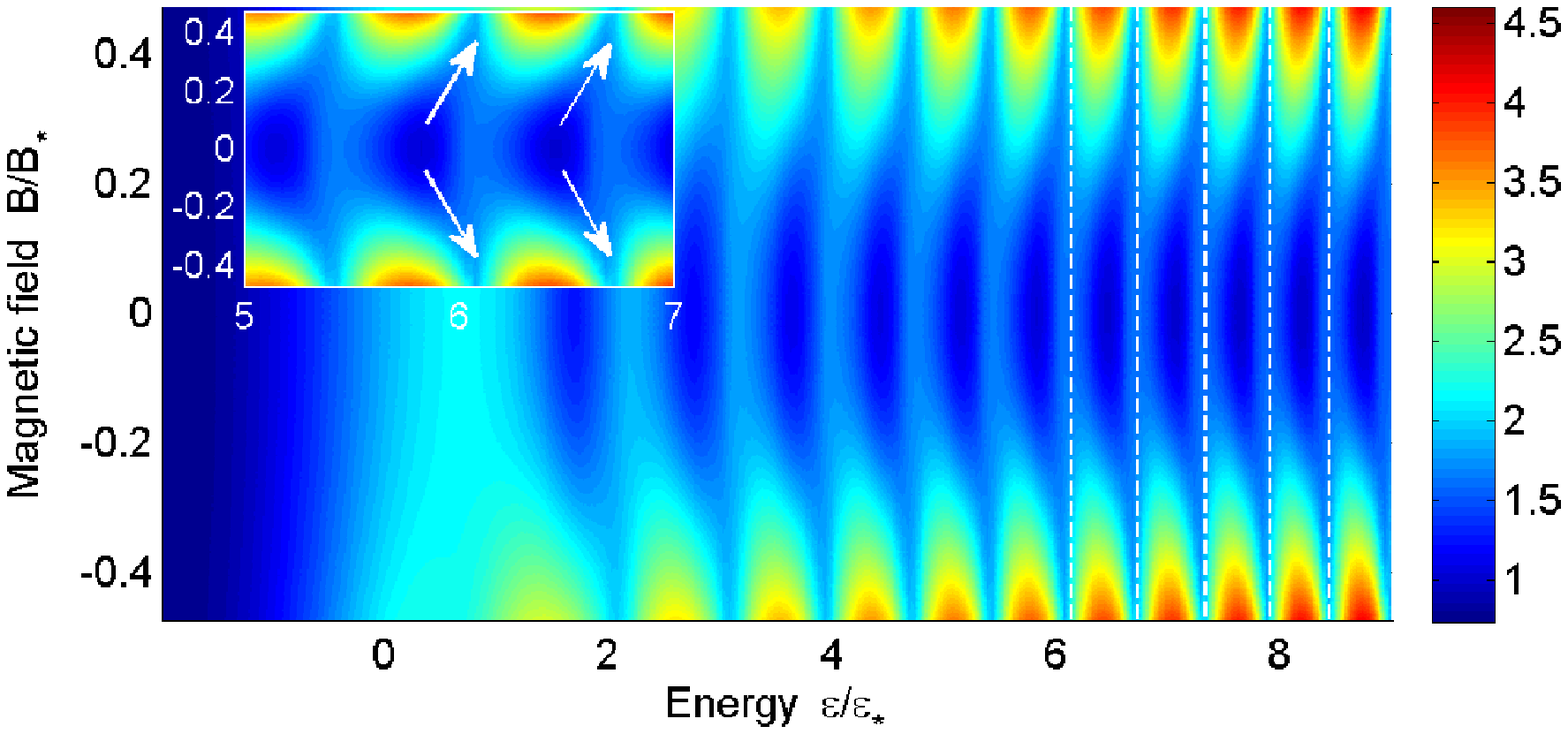}
\includegraphics[width=3.2in]{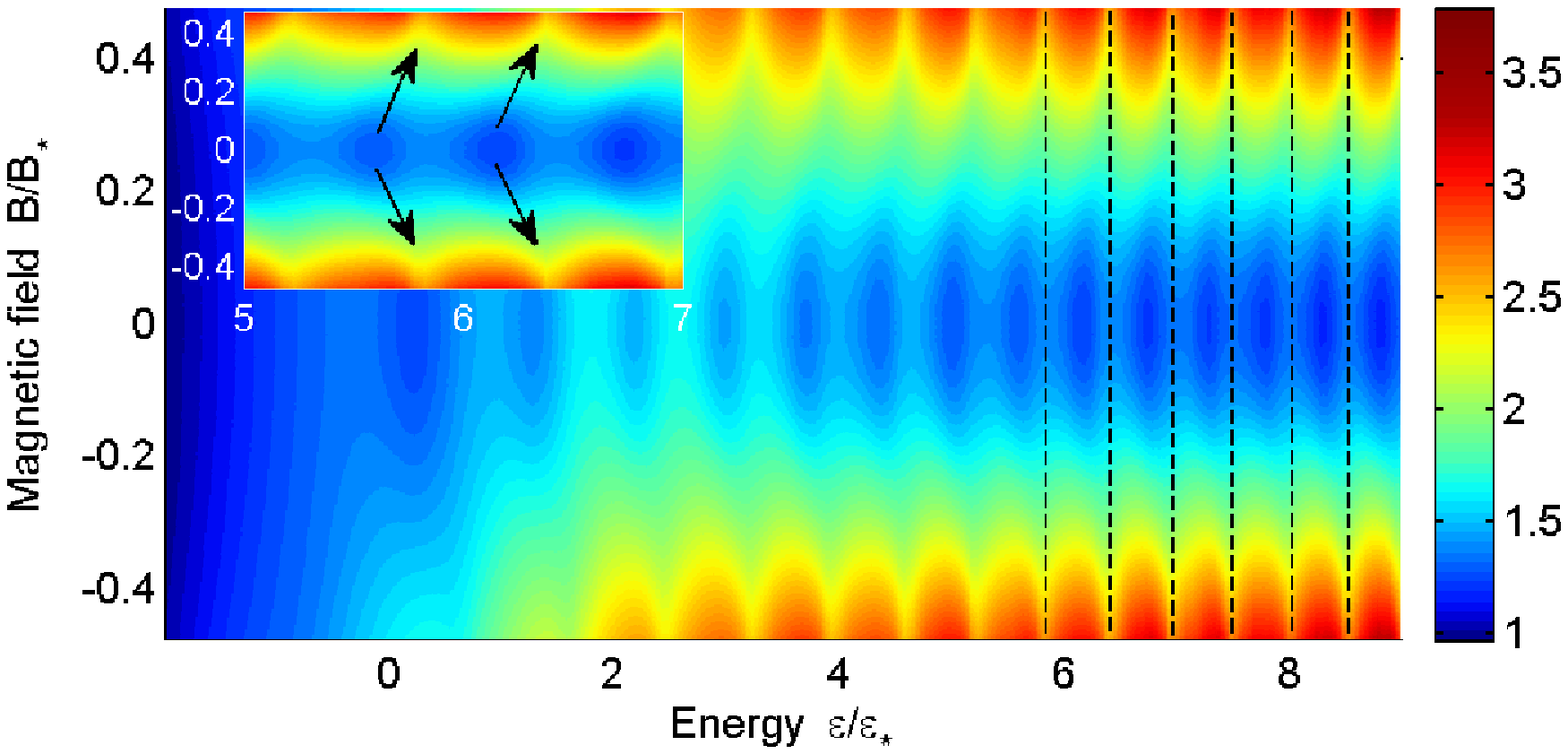}
\caption[t]{
Resistance (\ref{eq:resistance}) of the p-n-p structure as a function of magnetic field and energy. 
The quantity plotted is $\log R/R_*$, with $R_*=(x_*/W)h/4e^2$.
Note that the resistance minima at high $B$, marked by dashed lines, are shifted by half a period relative to those at zero $B$. In a close-up of a few fringes (insets) arrows indicate the  field-induced shift.
Lower panel illustrates the effect of averaging over smooth fluctuations, Eq.(\ref{eq:T_average}), with $\sum_m|a_m|=2.5\epsilon_*$, which suppress fringe contrast without affecting the phase shift. 
}
 \label{fig3}
\vspace{-5mm}
\end{figure}


These results can now be applied to analyze conductance and resistance, given by
\be\label{eq:resistance}
R(\epsilon)=G^{-1}
,\quad
G=
\frac{4e^2}{h} W\int^\infty_{-\infty}T(\epsilon,p_{y})\frac{dp_{y}}{2\pi}
\ee
where $W$ is the width of the p-n structure (see Fig.\ref{fig0}).
As illustrated in Fig.\ref{fig2}, the resistance exhibits fringes which obey the $n^{2/3}$ scaling, as expected from the phase dependence on the energy, Eq.(\ref{eq:WKB_theta}). Somewhat surprisingly, the integral over $p_{y}$ in (\ref{eq:resistance}) yields a fairly high fringe contrast in $R$. This is in agreement with the dispersion of resonances, Eq.(\ref{eq:WKB_theta}), being weaker than the dispersion of transmission $t_{1(2)}(p_{y})$. 


In the presence of magnetic field, alongside with the overall increase in resistance, we observe that the fringes shift up in energy by approximately half a period (see Fig.\ref{fig3}a). This shift, which is a direct consequence of the $\pi$-shift of the reflection phase discussed above, fully develops in the fields $B\sim 0.4 B_*$. 

The effect of large-scale potential fluctuations, either intrinsic \cite{Martin08} or induced by variable distance to gates, can be analyzed by averaging
the conductance in (\ref{eq:resistance}) over random offsets in $\epsilon$, taken as a sum of a few harmonics:
\be\label{eq:T_average}
\la T\ra=\oint T\lp \epsilon-\sum_m a_m\cos ( m\phi+\phi_m )\rp\frac{d\phi}{2\pi}
\ee
This simple model describes smooth inhomogeneity with the correlation length in excess of p-n interface separation $L$ but much
shorter than the structure width $W$.

The averaging procedure (\ref{eq:T_average}), applied to our numerical results, makes the fringes aperiodic and suppresses the contrast (see red dashed line in Fig.\ref{fig2}). However, the $\pi$ phase shift induced by magnetic field remains clearly discernible even for relatively strong fluctuations (see Fig.\ref{fig3}, where we use $\sum_m|a_m|=2.5\epsilon_*$ which is quite large compared to the fringe period of about $0.6\epsilon_*$).

At even stronger randomness, the FP transmission (\ref{eq:FPgeneral}) can be replaced by its phase-averaged value
\be\label{eq:T_ave}
\la T\ra_\theta \!\!= \frac{t_1t_2}{1-r_1r_2}\!\!=\frac1{e^{\lambda(p_{y}+eBx_\epsilon)^2}\!\!+e^{\lambda(p_{y}-eBx_\epsilon)^2}\!\!-1}
.
\ee
Plugged in (\ref{eq:resistance}), it yields magnetoresistance with characteristic $B\sim B_*$, identical to that discussed in \cite{Cheianov06}. The resulting exponential suppression of conductance of course holds only in the absence of short-range disorder.

Conspicuously, the resistance data \cite{Gorbachev08,Stander08} feature aperiodic oscillations in gate voltage, observed above the point where the sign of carriers beneath the top gate is reversed. This is the same region where strong FP fringes are expected for an ideal system. The energy scale of the oscillations reported in Ref.\cite{Gorbachev08}, converted from gate voltage using $\delta\epsilon/\delta V_{\rm tg}\approx \frac1{300}$, is about $\delta\epsilon\sim 30\,{\rm mV}$, which is only a few times larger than the period of $0.8\epsilon_*\approx 11\,{\rm mV}$ found above (Figs.\ref{fig2},\ref{fig3}). Could these oscillations, or those seen in \cite{Stander08}, be the FP fringes contaminated by disorder? Comparison to the behavior of the FP contrast in the presence of magnetic field, in particular to the $\pi$ phase shift (Fig.\ref{fig3}), may help to clarify this.

In summary, Fabry-P\'erot interference in the Klein scattering regime is found to be sensitive to the phase of the reflection amplitude that exhibits a jump by $\pi$ near zero incidence angle. This leads to half a period shift of interference fringes in the presence of a relatively weak magnetic field, a new effect that can help to identify the Klein phenomenon in graphene.

We thank A. K. Savchenko and D. Goldhaber-Gordon for useful discussions.
This work was supported by DOE Grant No. DE-FG02-06ER46313 (AS) and a National Science Foundation Graduate Research Fellowship (MR).

\end{document}